\documentclass[epjCONF]{svjour}
\usepackage{graphics}
\usepackage[varg]{txfonts} 
\usepackage[latin1]{inputenc}
\session-title{UHECR2012}
\begin{document}
\title{Theoretical Challenges in Acceleration and Transport of Ultra High Energy Cosmic Rays: A Review}
\author{Pasquale Blasi\inst{1,2}\fnmsep\thanks{\email{blasi@arcetri.astro.it}}}
\institute{$^{1}$INAF/Osservatorio Astrofisico di Arcetri\\
$^{2}$INFN/Laboratori Nazionali del Gran Sasso}
\abstract{
The wealth of data collected in the last few years thanks to the Pierre Auger Observatory and recently to the Telescope Array made the problem of the origin of ultra high energy cosmic rays a genuinely experimental/observational one. The apparently contradictory results provided by these experiments in terms of spectrum, chemical composition and anisotropies do not allow to reach any final conclusions as yet. Here I will discuss some of the theoretical challenges imposed by these data: in particular I will discuss some issues related to the transition from Galactic to extragalactic cosmic rays and how the different models confront our understanding of Galactic cosmic rays in terms of supernova remnants paradigm. I will also discuss the status of theories aiming at describing acceleration of cosmic rays to the highest energies in relativistic shocks and unipolar inductors. 
} 
\maketitle
\section{Introduction}
\label{intro}
The origin of the highest energy cosmic rays has been challenging our models of particle acceleration and propagation for a long time. The detection of events with energy $\gtrsim 10^{20}$ eV pioneered by John Linsley \cite{linsley} turned out to be hard to reconcile with the discovery of the cosmic microwave radiation fossile of the big bang: soon after the discover of such background radiation the idea was put forward that the spectrum of ultra high energy cosmic rays (UHECRs) should end at $\sim 10^{20}$ eV as a result of the onset of photopion production on the cosmic microwave radiation, the so called GZK feature \cite{gzk1,gzk2}. The search for the GZK feature in the data has had many twists and turns: about a decade ago it appeared that the spectrum, as measured by the AGASA experiment extended above $\sim 10^{20}$ eV \cite{agasaGZK}; many models were proposed to explain how to explain this, from the effect of local sources to top-down models and violations of fundamental laws of Physics, such as the principle of Lorentz invariance. On the other hand, the HiRes experiment, with a similar exposure claimed the detection of the GZK feature \cite{hiresGZK}. In fact, both experiments had very poor statistics of events and both the absence or the presence of the GZK feature were rather weakly established \cite{demarco}. More recently the HiRes experiment reiterated the claim for detection of the GZK feature \cite{hires} and the existence of the feature has finally been confirmed by the Pierre Auger Observatory first \cite{augerGZK} and by the Telescope Array \cite{TA-GZK} later. This did not close the debate on the GZK feature: in ref. \cite{grigo} it was first proposed that $E_{1/2}$, the energy at which the modification factor in the flux of UHECRs is reduced to 1/2 of the value inferred from the lower energy extrapolation, can be used as a powerful indicator of the presence of the GZK feature and its association to photopion production. The HiRes collaboration measured this quantity \cite{hires} and found its value to be $10^{19.73\pm 0.07}$ eV, in perfect agreement with the predicted value of $10^{19.72}$ eV \cite{grigo}. This agreement reinforces the statement that the flux reduction observed by HiRes may in fact be the GZK feature. However, the Pierre Auger Collaboration measures a flux reduction that starts at somewhat lower energies. Moreover, while the chemical composition measured by HiRes (and more recently confirmed by Telescope Array) by using the elongation rate is consistent with a proton dominated composition at energies above $10^{18}$ eV, the Pierre Auger Observatory measures a gradual change in composition from light to heavy moving toward higher energies, and basically dominated by Iron at $\sim 50 EeV$ \cite{augerchem}. Since the GZK feature is typical of the proton spectrum, the flux suppression measured by Auger would therefore not be the GZK suppression but rather some signature of nuclear photodisintegration, or even an intrinsic cutoff in the acceleration spectrum. Last but not least, the Pierre Auger Observatory found correlation of the arrival direction of UHECRs above 57 EeV of energy with the position in the sky of AGN in selected catalogs \cite{science}. The statistical significance of the claim has been decreasing with time and appears to have reached a (stable) significance of $\sim 2\sigma$ at present \cite{correlation}. The correlation suggests that the chemical composition is light in order to avoid excessive deflection in the magnetic field of the Galaxy. 

What to explain then? Are there really particles accelerated in Nature to energy $\gg 10^{20}$ eV or we are detecting an intrinsic cutoff in the accelerated spectrum? Is the observed flux suppression due to photopion production of protons or photodisintegration of nuclei? Should the accelerator energize mainly protons or mainly heavy nuclei? Is the correlation real or is it just suggesting what can naturally be expected, namely that UHECRs mainly come from the regions where matter is concentrated? The ambiguous answers to these questions clearly indicate that at the present time most of the issues in looking for the origin of UHECRs are related to observational/experimental matters. The most important source of uncertainly is the measurement of chemical composition, both in the so-called transition region and at the highest energies. Several dedicated efforts are being made to understand composition in the energy region $10^{18}-10^{19}$ eV (from KASCADE-GRANDE to the Auger enhancements and the Telescope Array efforts). At the highest energies the issue is more subtle: the statistics of events does not appear to be the problem any more; in principle the statistical error bars on the elongation rate and its fluctuations are small enough to allow for the discrimination between light and heavy composition. The question is whether either a poor knowledge of shower physics and/or unknown systematics and/or different cuts in the data may affect the conclusions and perhaps explain the apparently contradictory conclusions reached by different experiments. 

The experimental challenges in the field of UHECRs will be reviewed in the presentation of P. Privitera.
Here I will limit myself with discussing some theoretical aspects that appear to be relevant and open, even independent of the confusing inputs we are receiving from observations. More specifically in \S \ref{sec:transition} I will briefly discuss the issue of the end of Galactic cosmic rays and beginning of extragalactic cosmic rays in the light of recent developments concerning the so-called supernova remnant paradigm for Galactic CR. A more dedicated discussion of the transition region can be found in the presentations of V. Berezinsky and R. Aloisio (these proceedings). In \S \ref{sec:simple} I will summarize some solid facts that can be stated on UHECR accelerators (almost) without knowing what they are. In \S \ref{sec:shock} I will discuss some open issues concerning  acceleration at shocks and connections with Physics of large scale structures, AGN and gamma ray bursts (GRBs). In \S \ref{sec:ns} I will summarize some recent developments concerning rapidly rotating neutron stars as possible sources of UHECRs. In \S \ref{sec:cascade} I will briefly discuss the possibility to receive high energy gamma rays from distant sources, due to the electromagnetic cascade initiated by UHECRs. I will summarize in \S \ref{sec:summary}.

\section{The transition from Galactic to Extra-Galactic Cosmic Rays}
\label{sec:transition}

The quest for the origin of UHECRs is tightly related to the issue of the transition from Galactic cosmic rays to CRs generated in extragalactic sources. For long time it has been taken for granted that the ankle in the CR spectrum, at $\sim 10^{19}$ eV, is the spectral signature of the transition from a steep Galactic spectrum to a flatter extragalactic spectrum. The situation has however changed and the nature of the ankle questioned as a consequence of three developments: a) in Ref. \cite{dip} the authors noticed that Bethe-Heitler pair production leaves a distinct feature in the spectrum of CRs propagating on cosmological scales. The feature takes the form of a dip whose shape fits very well the observed modification factor for all experiments, with the possible exception of the one measured by the Pierre Auger Observatory. In this model CRs with energy $\ge 1$ EeV are of extragalactic origin, and the transition occurs at the second knee. b) In Ref. \cite{mix} the authors discussed the possibility that UHECRs may be nuclei with a mixed chemical composition. In this model the Galactic component of CRs ends at energy $\sim 2$ EeV. The so-called disappointing model introduced in Ref. \cite{disapp} is a special case of the mixed composition model, in which the maximum energy of protons is relatively low, $\sim 4\times 10^{18}$ eV, and the iron spectrum extends to $\sim 10^{20}$ eV. The model is {\it disappointing} in that the flux suppression at $\sim 10^{20}$ eV is not the GZK feature but rather the intrinsic cutoff in the source spectrum and no correlation is expected because of the heavy composition at the highest energies. 

\begin{figure}
\begin{center}
\resizebox{.8\columnwidth}{!}{%
  \includegraphics{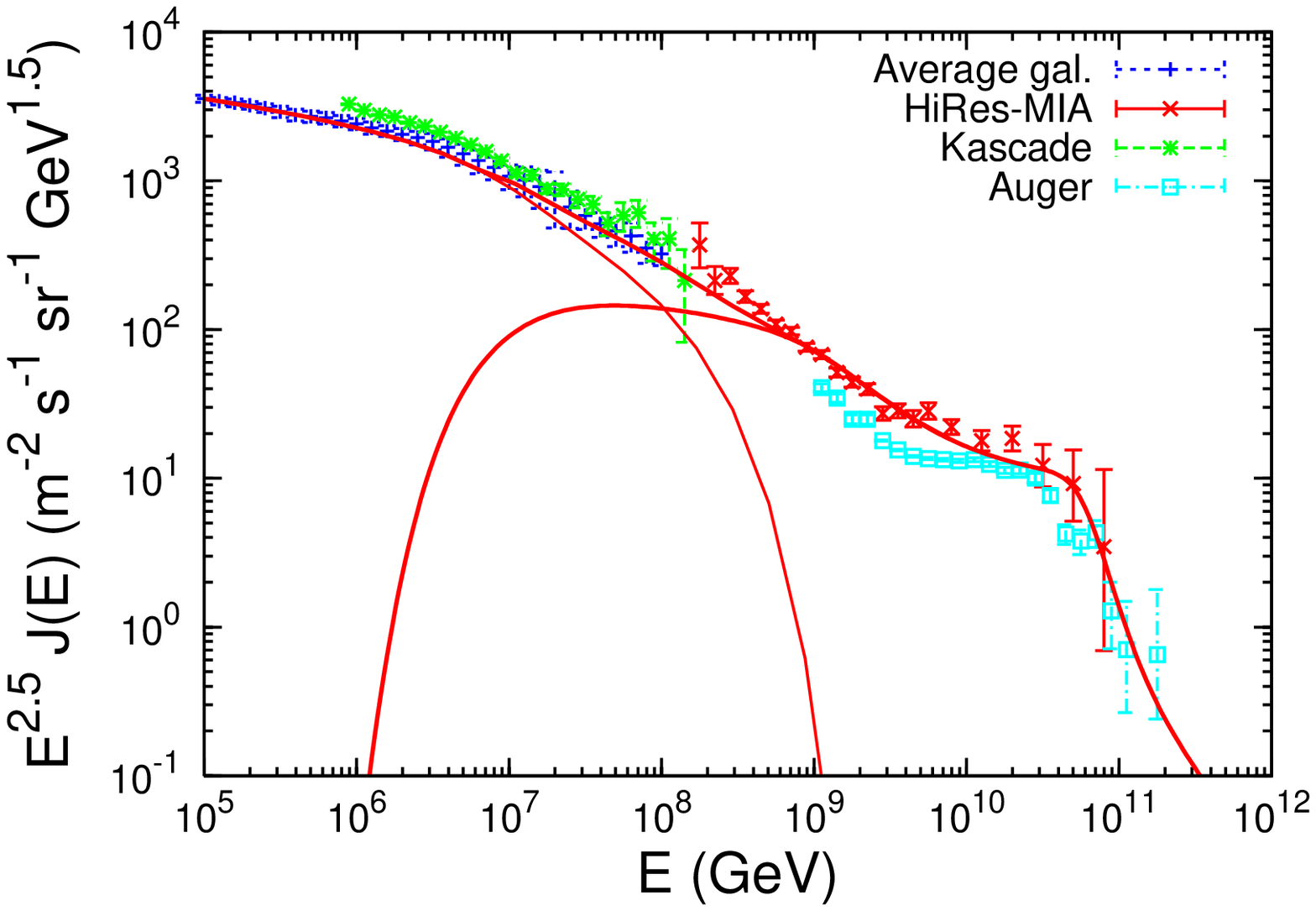} 
  \includegraphics{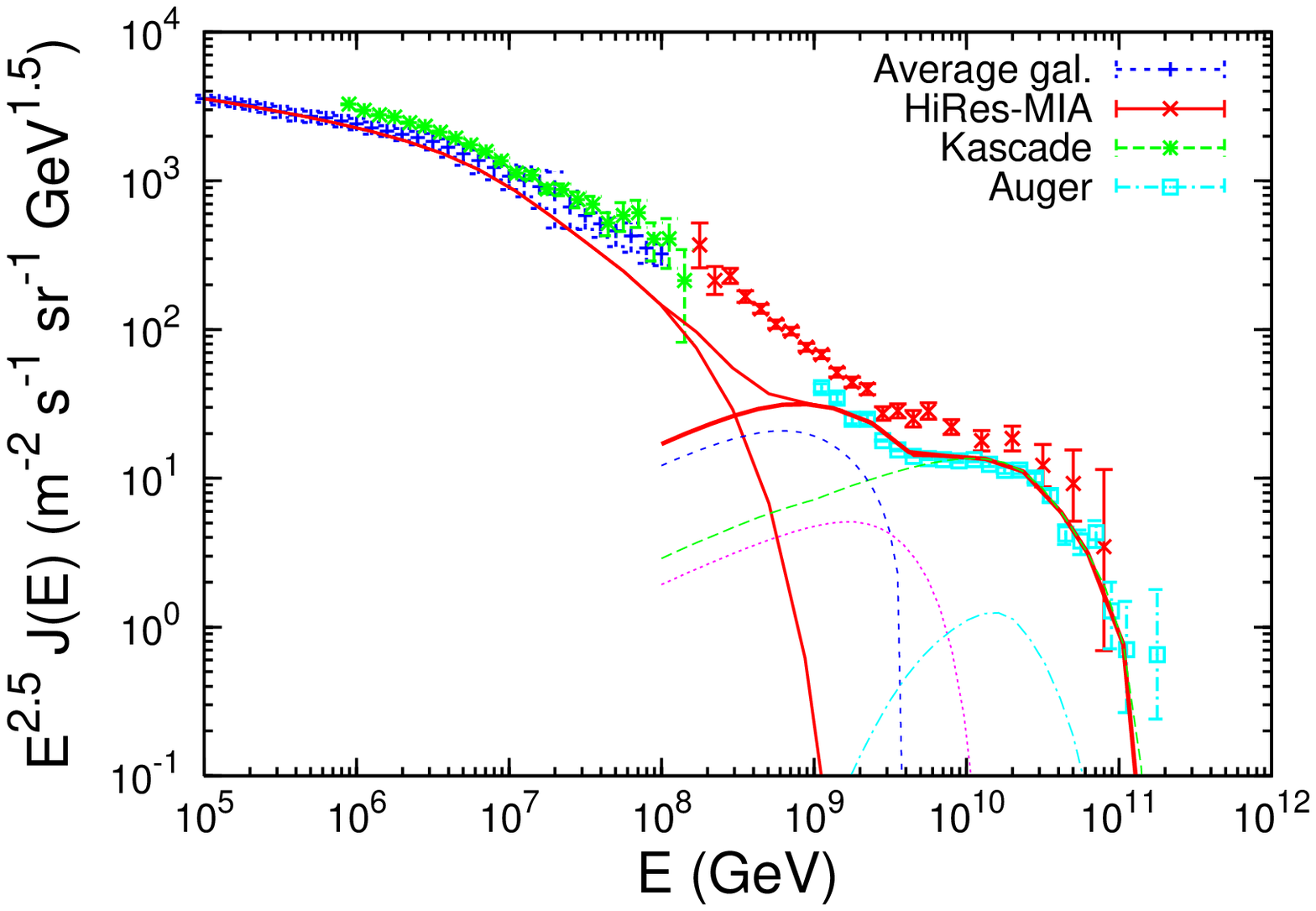} 
  }
\caption{Left Panel: Spectrum of CRs in the dip model overlapped to the Galactic CR flux. Right Panel: Spectrum of CRs in the disappointing model with the Galactic CR flux as in the left panel.}
\label{fig:spectra}       
\end{center}
\end{figure}

Both these models (dip and mixed composition) lead us to expect that Galactic CRs may end in the EeV region (with a composition dominated by heavy nuclei) rather than at the ankle. This conclusion also appears to be supported by a recent investigation of cosmic ray anisotropies \cite{giacinti}. 

c) This conclusion turns out to be exciting since the supernova remnant (SNR) paradigm which has been developed in the last decade or so in its modern form, based on the non-linear theory of particle acceleration at supernova shocks, suggests that CRs accelerated in SNRs may reach maximum rigidity $\sim$ a few $10^{6}$ GV. For an iron nucleus this implies a cutoff at $\sim 10^{17}$ eV. Although more rare types of supernovae may possibly reach higher energies than these, this possibility appears to be purely speculative at this time and not required based on available data \cite{amato1}.

The spectrum of cosmic rays is shown in Fig. \ref{fig:spectra} for the dip model (left panel) and for the mixed composition model in its disappointing model configuration (right panel), as calculated by \cite{aloprep} (see also the paper of R. Aloisio in these proceedings \cite{bob}). In these calculations the spectrum of Galactic CRs was taken from \cite{amato1}. In the right panel, the different lines represent the fluxes of different chemicals, the solid line representing the total flux. One can clearly see that at high energy the chemical composition is dominated by heavy nuclei. The transition region is better described by the dip model. 

\begin{figure}
\begin{center}
\resizebox{.8\columnwidth}{!}{%
  \includegraphics{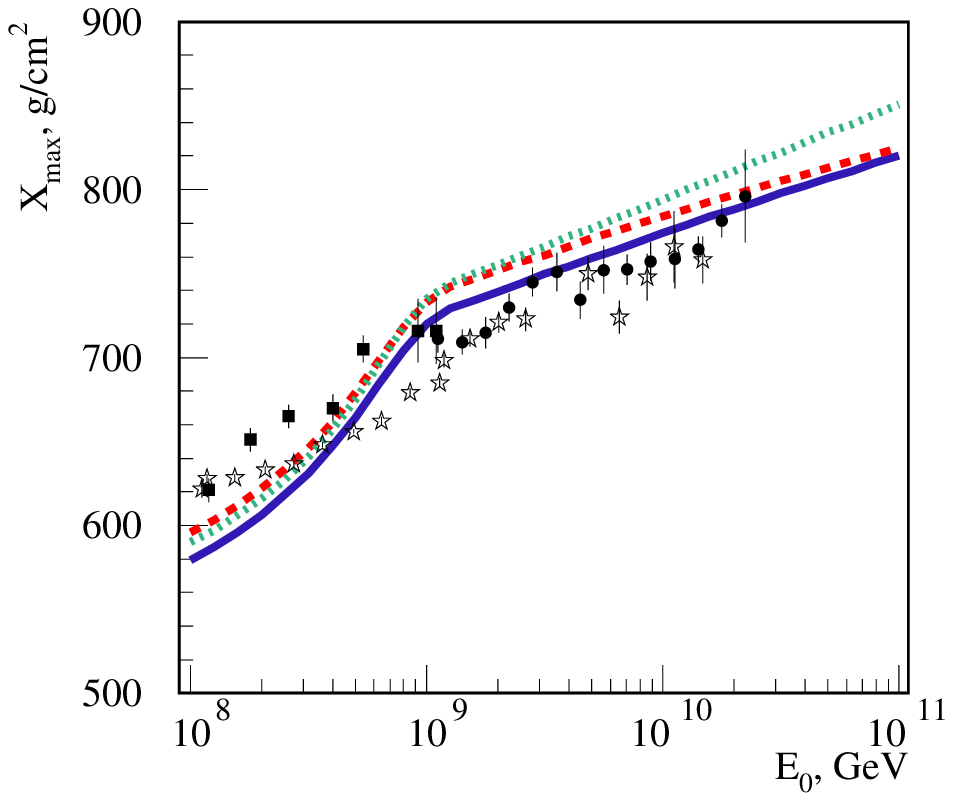} 
  \includegraphics{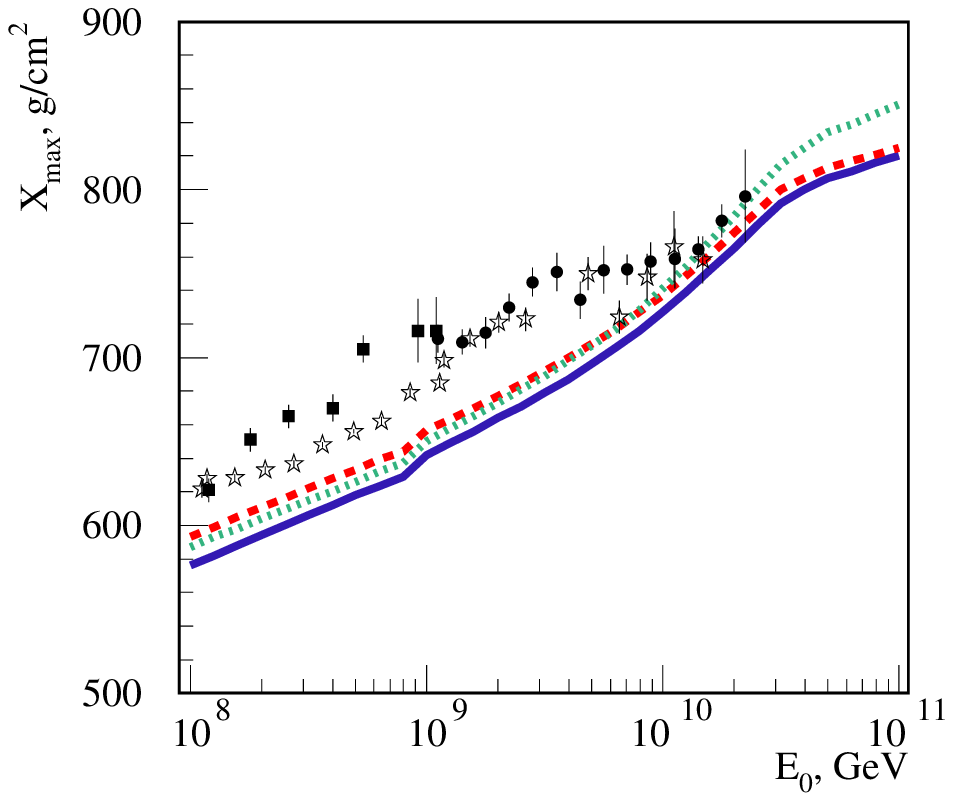} 
  }
\caption{Left Panel: Elongation rate for the dip model (lines) compared data from different experiments (see \cite{uscomp} for details). Right Panel: Elongation rate for the ankle model. The solid, dash and dotted lines refer to QGSJET, QGSJET-II and SYBILL for the development of the showers.}
\label{fig:comp}       
\end{center}
\end{figure}

The main discrimination among dip model, mixed composition model and ankle model is based upon the measurement of the chemical composition, especially in the transition region \cite{uscomp}. In the dip model the elongation rate is expected to show a sharp transition from a heavy Galactic composition at energy below the second knee to a light composition at $E>10^{18}$ eV (see \cite{usdip}) as illustrated in Fig. \ref{fig:comp} (left panel). The solid, dash and dotted lines refer to QGSJET, QGSJET-II and SYBILL for the development of the showers. The transition is predicted to be concluded at $10^{18}$ eV, where the composition is completely dominated by protons. The dip model works well provided the abundance of Helium in the primaries is smaller than $\sim 10\%$ in flux. 

In the mixed composition model the extragalactic CRs are made of a mix of different elements, and the flux at Earth is the result of a complex chain of interactions: photodisintegration of nuclei leads to lighter composition even when starting with pure Iron at the source. One can change the injection spectra and the composition at the source to match the observed composition and spectra, although in general it is hard to explain a transition to heavy composition at ultra high energies, as it is observed by Auger (see \cite{bob}). In the case of the mixed composition model the elongation rate shows a gradual transition from heavy galactic CRs to somewhat lighter extragalactic CRs.The so-called disappointing model reproduces the heavy composition of Auger at high energy by construction. 

In the traditional ankle model the transition reflects in a gradual change from an Iron dominated composition of Galactic CRs that extend to $\sim 10^{19}$ eV to a pure proton composition, reached at energies $\sim 5\times 10^{19}$ eV, so that in this case only a small region of energies is filled by the extragalactic CR component. This model does not fit the elongation rate as observed in any of the current experiments  (see right panel of Fig. \ref{fig:comp}) and is not immediately compatible with the SNR paradigm for the origin of Galactic CRs. 

As discussed above, both the dip model and the mixed composition model require that galactic CRs have a cutoff at energies between $\sim 10^{17}$ eV (for the dip model) and $\sim 10^{18}$ eV (for a mixed composition). It is therefore important to check the consistency of this prediction with the expectations based on the SNR paradigm. 

The most convincing evidence that at least some SNRs accelerate CRs in the Galaxy is provided by 1) gamma ray emission in selected SNRs; 2) morphology and spectrum of the X-ray emission; 3) anomalous widths of Balmer lines in selected SNRs with Balmer dominated shocks. Below, I will briefly discuss these three points. 

Particle acceleration in SNRs is described using the non-linear theory of particle acceleration (see \cite{maldrury} for a review of the non linear theory of diffusive shock acceleration - NLDSA), which takes into account the dynamical reaction of the accelerated particles on the shock. In more modern versions \cite{amato2006} the phenomenon of cosmic ray driven magnetic field amplification and dynamical reaction of the amplified fields on the shock \cite{capri} are also taken into account. The phenomenon of magnetic field amplification induced by streaming instability of CRs is central to the issue of the origin of CRs in SNRs, in that only this process can lead to maximum energies of relevance for CRs, at least in parallel shocks. For perpendicular shocks, which may occur when a supernova explosion occurs in the wind of the presupernova star, one may argue in favor of shock drift acceleration, which reduces the acceleration time \cite{joki}.

Magnetic fields in excess of the ones usually observed in the interstellar medium (ISM) have been inferred in the last decade from numerous SNRs from the effects they cause on the morphology of the X-ray emitting region: the observed emission is synchrotron radiation from highly relativistic non thermal electrons, accelerated at the shock; the width of the emission region is limited by synchrotron losses and the comparison of predicted and observed morphology leads to infer magnetic fields in the range $100-1000 \mu G$ behind the shock. This finding has been viewed as a possible indication that accelerated particles amplify the magnetic field as was long expected (see for instance \cite{bell78}). The streaming instability excited by CRs can proceed in a resonant \cite{achterberg} or non resonant \cite{bell2004} way. The two branches of the instability may coexist and prevail one on the other in different stages of the SNR evolution \cite{kinetic2009}. The non resonant branch may grow faster for young SNRs, but it is not clear if this reflects in shorter acceleration times of the particles since the growing modes occur on spatial scales much shorter than the Larmor radius of particles. In the following I will concentrate on the resonant mode of the instability, which leads to magnetic field values which are in good agreement with the observed ones under reasonable assumptions. It is however fair to say that in both cases the value of the magnetic field reached after the instability gets saturated is poorly known from theory, and is mostly based either upon extrapolations of quasi-linear theory to regimes where it should not be applied, or on numerical models with plenty of simplifying assumptions necessary to make the computation possible.  

\begin{figure}
\begin{center}
\resizebox{.9\columnwidth}{!}{%
  \includegraphics{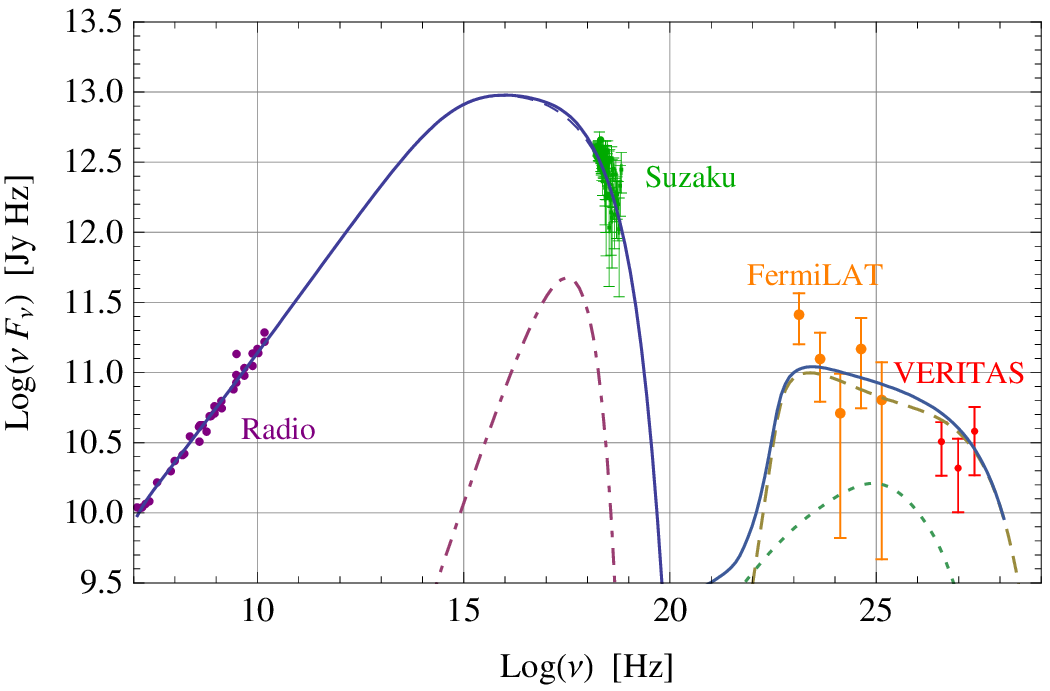} 
  \includegraphics{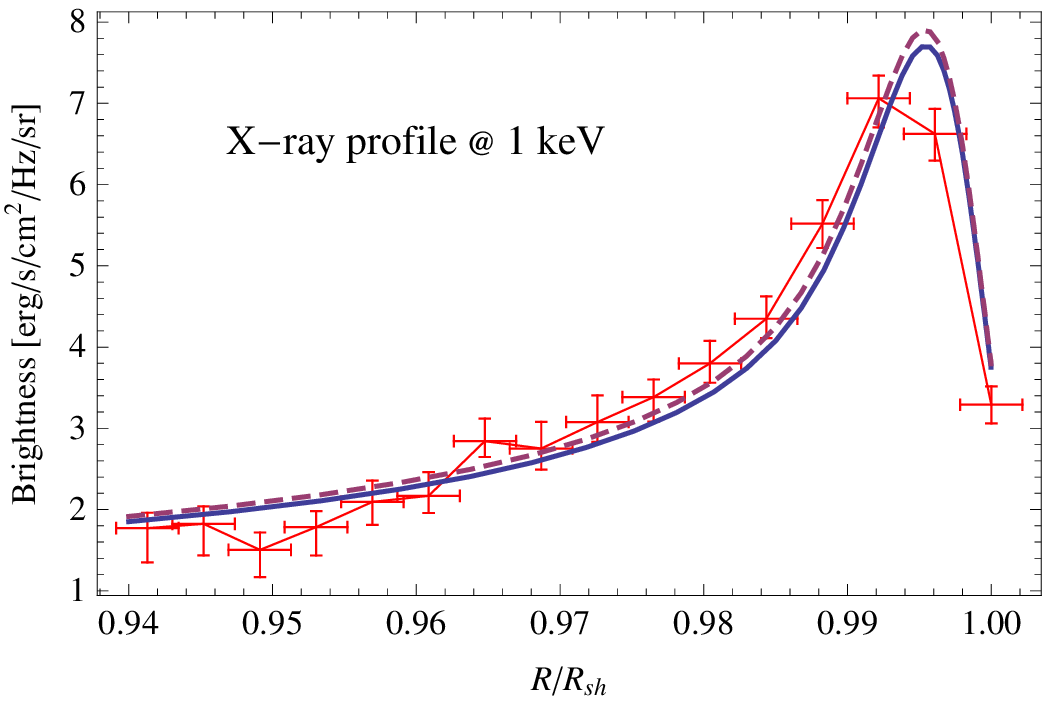} 
  }
\caption{Left Panel: Spatially integrated spectral energy distribution of Tycho \cite{morlino}. The curves show the calculated multifrequency spectrum from the radio to the gamma ray band. Right panel: Projected X-ray emission compared with the data from \cite{cassam}. The solid line shows the projected brightness of synchrotron emission convolved with the Chandra point spread function.
}
\label{fig:tycho}       
\end{center}
\end{figure}

There are several cases of SNRs close to molecular clouds where gamma ray emission has been observed and can be attributed to pion production, thereby indicating that some level of CR acceleration takes place. However these remnants are hardly the ones responsible for the acceleration of the bulk of Galactic CRs, since they are typically very old remnants. Moreover, the issue remains of whether the gamma ray emission from the SNR-cloud association tells us more about CR acceleration or CR propagation. At present the only young remnant for which a clear modeling of the multifrequency emission has been possible and strong evidence for gamma rays of hadronic origin could be found is the case of Tycho \cite{morlino}. Both gamma rays in the GeV \cite{tychofermi} and TeV \cite{tychoveritas} energy range have been detected, spectrum and morphology of the radio and X-ray emission exist and the overall appearance of the remnant is quite close to spherical, thereby suggesting that probably the environment in which the supernova occurred was not very complex. This is also consistent with the fact that the supernova event associated with Tycho is of Type Ia. The multifrequency spectrum predicted for Tycho is illustrated in Fig. \ref{fig:tycho} (left panel) where data points refer to the results of observations in the relevant bands and the lines are the predictions of Ref. \cite{morlino}. In the right panel of Fig. \ref{fig:tycho} the expected brightness profile of non-thermal X-rays (line) is compared with Chandra data. If the X-ray rims are due to synchrotron loss dominated propagation, then the magnetic field in the shock region can be estimated to be $\sim 300\mu G$. Notice that the gamma ray spectrum cuts off at $\sim 50$ TeV, suggesting that at the Tycho age the maximum energy of accelerated CRs should be $\sim 5\times 10^{14}$ eV. The maximum energy is a function of time and reaches a maximum at the beginning of the Sedov-Taylor phase. 

It is worth noticing that gamma ray emission likely of hadronic origin has also been detected from Cas A \cite{casA}, which is a very young remnant. Modeling of this remnant is however very complex: Cas A was a core collapse supernova and there are indications that most of the particles responsible for the radio emission are being accelerated at the reverse shock. This complexity is typical of core collapse supernovae: they often explode in the wind of the presupernova star, where the circumstellar medium is on average very underdense, and the background magnetic field is expected to be mostly azimuthal. This scenario complicates the description of the particle acceleration process and in addition is likely to imply a suppression of gamma ray emission (because of the low density in the bubble), unless the remnant is extremely young (see for instance \cite{caprioli} for a discussion of the effect of the red giant wind). This is rather disappointing in that the most frequent supernovae (Type II) and the biggest contributors to CR acceleration might well be very weak gamma ray sources during the phases of their evolution in which the ejecta expand in the bubble excavated by the presupernova wind. It is however important to realize that acceleration can proceed efficiently even if weak gamma ray emission is produced. 

The less frequent SNe Type Ia might be better targets for gamma ray observations, and they are easier to model because of the simpler environment in which they typically explode. It is also worth recalling that although it is difficult to have a proof that gamma rays have been produced in pion decays, there is unambiguous evidence that electrons are accelerated up to energy in the 10-100 TeV range. It is conceivable to assume that nuclei are accelerated too, but electrons simply have much larger radiative efficiency. A sort of proof that this may be the case comes from the fact that even in the cases where the gamma ray emission is best modeled with leptonic models \cite{elli1713}, there is evidence for magnetic field amplification at the shock (see for instance the case of 
RX J1713.73946). 

In the last few years another method has provided independent evidence for particle acceleration in SNR, namely the observation of the widths of Balmer lines  from Balmer dominated shocks (see \cite{heng} for a review). The underlying Physics is very simple: when a collisionless shock moves in a partially ionized medium, only the ionized component is shocked, while the neutral gas crosses the shock undisturbed. Hydrogen atoms that suffer a charge exchange reaction downstream of the shock acquire a temperature that leads to a broad Balmer line. A narrow component of the Balmer line is produced by the excitation of those neutrals that did not suffer charge exchange and retain the original temperature $T\leq 10^{4}$ K. The measurement of the width of the broad Balmer line allows us to infer the ion temperature downstream. For a shock that accelerates CRs efficiently one expects, based on NLDSA, that the temperature of ions downstream is lower that in the absence of CRs, thereby leading to a narrower broad Balmer line. On the other hand, in NLDSA a CR induced precursor is induced upstream of the shock, so that charge exchange between warmer ions and cold neutrals can take place. In this case the narrow Balmer line becomes broader. Both these anomalies in the widths of the narrow and broad Balmer lines are diagnostic tools that can be used to measure the CR acceleration efficiency. This exercise has been applied to RCW 86 \cite{vink} where the authors use the broad line to infer a CR acceleration efficiency of $> 50\%$. In \cite{lee} the measurement of the Balmer line width in Tycho also led to conclude that observations are consistent with efficient particle acceleration taking place. 
These conclusions are however in general model dependent, and it can hardly be otherwise since a full theory of NLDSA in partially ionized media is only now being developed. In Ref. \cite{vinktheory} the authors made an attempt to model the effect of the CR induced precursor but the structure of the shock is not calculated self-consistently. In \cite{usneutral} the authors discuss several new effects associated with the presence of neutrals that potentially affect the shock structure in an important way. 

All these methods show evidence for efficient particle acceleration in SNRs. Unfortunately, from gamma ray observations, the evidence for acceleration of hadrons is so far limited to the Tycho SNR and to a few old remnants located close to molecular clouds. As discussed above this is however not that surprising. In the case of Tycho the maximum energy of accelerated protons is inferred to be $\gtrsim 500$ TeV, in agreement with the value inferred using Bohm diffusion of protons in the amplified magnetic field measured from the morphology of the X-ray emission. Somewhat higher energies, even closer to the knee, can be obtained for SNRs closer to the beginning of the Sedov phase. Clearly none of these arguments rules out the possibility that some rare type of supernova explosion may lead to much larger maximum energy of accelerated particles, but there is certainly no evidence that this may happen and there is no need to postulate that it may happen. At present one can safely conclude that the supernova paradigm based on NLDSA and current observations suggest that SNR can accelerate CRs up to rigidities in the range $\sim 10^{6}-10^{7}$ GV, consistent with what is required to Galactic CRs in the context of the dip and mixed composition models. 

\section{What can we say about the sources without knowing what they are?}
\label{sec:simple}

We do not know what the sources of UHECRs are. We do not know if the acceleration region is moving relativistically or not. In this situation it becomes important to make an attempt at outlining the basic requirements for the accelerator. Below I will sketch a line of reasoning that resembles the one initially put forward in Ref. \cite{waxman}. Let us start from the non relativistic case: let the size of the acceleration region be $R$ and let the magnetic field be $B$. The condition required for the accelerator to reach particle energy $E$ is that $r_{L}(E)<R$, where $r_{L}(E)=E/(ZeB)$ is the Larmor radius of the particle with charge $Ze$. This condition can be transformed to a condition on the magnetic field:
\begin{equation}
B>\frac{E(eV)}{300 Z R} \to \epsilon_{B}=\frac{B^{2}}{4\pi} > \frac{E(eV)^{2}}{4 \pi(300 Z R)^{2}}. 
\end{equation}
The magnetic energy density has to be smaller than the total ram pressure $\epsilon_{B} < \rho V^{2}$, which translates to a lower limit on the luminosity of the source:
\begin{equation}
{\cal L} = 4\pi R^{2} \frac{1}{2} \rho V^{2} V > 2\pi R^{2} V \epsilon_{B} = 1.6 \times 10^{45} Z^{-2} \left(\frac{E}{10^{20} eV}\right)^{2} \beta ~ \rm erg/s,
\end{equation}
where $\beta=V/c$ and $V$ is the velocity of the acceleration region. One thing to notice is the strong dependence of this lower limit on the charge of the nucleus. For an iron nucleus ($Z=26$) the constraint becomes much less severe than for protons. One should also notice that there are caveats in the way the bound is obtained. For instance one could apply the bound to the specific case of acceleration of diffusive type (such as at a shock) with Bohm diffusion coefficient $D(E)=(1/3) r_{L} c$. In this case the condition of the Larmor radius becomes:
\begin{equation}
\frac{1}{3}\frac{r_{L}c}{V} < \xi R, ~~~\xi<1.
\end{equation}
In terms of magnetic field this reads:
\begin{equation}
\epsilon_{B}>9.8\times 10^{-8} \frac{E(eV)^{2}}{Z^{2} \beta^{2} \xi^{2} R^{2}},
\end{equation}
and in terms of luminosity:
\begin{equation}
{\cal L} > 1.8\times 10^{46} Z^{-2} \left(\frac{E}{10^{20} eV}\right)^{2}\left(\frac{\xi}{0.1}\right)^{-2} \beta^{-1} ~ \rm erg/s.
\end{equation}
Although the scaling with charge is the same as before, the scaling with $\beta$ is now very different and depends on the parameter $\xi$, the fraction of the size $R$ upstream of the shock where escape of particles may occur. 

The relativistic case is somewhat less ambiguous: let us consider a region of size $R$ and a particle with energy $E$, both in the laboratory frame. The accelerator is moving with Lorentz factor $\Gamma\gg 1$. Let us indicate with a prime all quantities measured in the comoving frame. For instance $E'=E/\Gamma$. The condition for acceleration to energy $E$ in the comoving frame reads 
\begin{equation}
T_{acc}=\frac{2\pi r_{L}(E')}{c} < T_{dyn} = \frac{R'}{c}=\frac{R}{\Gamma c},
\end{equation}
where $T_{acc}$ is the acceleration time and $T_{dyn}$ is the dynamical time, and we used Lorentz contraction of length scales. Taking into account that 
$r_{L}(E') = E'/(Z e B')$ the limit above translates to:
\begin{equation}
\epsilon_{B'}=\frac{B'^{2}}{4\pi} > \left( \frac{E(eV)}{300 Z R}\right)^{2}
\end{equation}
and since the energy density transforms as $\Gamma^{2}$ the limit on luminosity becomes:
\begin{equation}
{\cal L} > 4\pi R^{2} c \Gamma^{2} \epsilon_{B'} \approx 10^{47} \Gamma^{2} Z^{-2} \left(\frac{E}{10^{20} eV}\right)^{2}~\rm erg/s.
\end{equation}

These constraints on luminosities do not leave much room for the possible sources of UHECRs: if non relativistic, shocks associated with large scale structure formation and possibly large radio lobes in powerful radio galaxies might accelerate particles to $\sim (5-8)\times 10^{19}$ eV (see below). If relativistic, very bright AGN and GRBs remain allowed (although the possibility of nuclei makes these constraints a bit more relaxed). 

The energy input per unit volume is another quantity that we can write without knowing much about the sources of UHECRs. The flux of UHECR with energy $E$ can be roughly written as:
\begin{equation}
F(E) \approx \frac{c}{4\pi} \dot n(E) \tau_{loss}(E),
\end{equation}

where $\dot n$ is the injection rate of particles with energy $E$ and $\tau_{loss}(E)$ is the loss length of particles with energy $E$. Comparing this expression with the observed flux at, say, $10^{19}$ eV, one immediately deduces an energy input in the form of particles with $E>10^{19}$ eV of $\sim 3\times 10^{45}~\rm erg/Mpc^{3}/yr$. Unfortunately this number does not tell us much in the absence of clear evidence for small scale anisotropies. If no small scale anisotropy is observed at energies $\sim 10^{20}$ eV, one could impose a lower limit on the source density of continuous sources of $N_{source}>10^{-4}\rm Mpc^{-3}$. 

\section{Basic issues involved in the Physics of particle acceleration at shocks}
\label{sec:shock}

The simple arguments illustrated in the previous section do not explain how particles can be energized to the observed ultra high energies. In the following I will briefly discuss some instances of newtonian and relativistic shock waves as possible accelerators. 

\subsection{Non relativistic shocks and large scale structures}

The principles of particle acceleration at non relativistic shocks are relatively well understood. A charged particle that diffuses through a collisionless shock front that moves with velocity $V_{s}$ gains energy at each crossing, so that $\Delta E/E=\frac{4}{3c} (V_{sh}-V_{sh}/4)=V_{sh}/c$, where we assumed that the shock is strong and therefore the compression factor is 4. In most cases of interest the assumption of strong shock works reasonably well. The spectrum of accelerated particles is a power law in momentum $N(p)\propto p^{-s}$ with $s=2$ for a strong shock. Here the particle number is normalized so that $N(p)dp$ is the number of particles in the energy bin $dp$ around momentum $p$. These results apply to the so-called test-particle theory, namely they are based on the assumption that the dynamical reaction of accelerated particles on the shock can be neglected. For most cases this is not a good approximation, and several effects arise as a result of the inclusion of non-linearities (for instance magnetic field amplification, which in turn speeds up the acceleration process, and spectral concavity). 

An easy estimate of the acceleration time can be obtained by assuming that diffusion occurs in the Bohm limit $D(E)=(1/3) r_{L}(E) c$, which minimizes the acceleration time (maximizes the highest energy that can be achieved). The acceleration time should be compared with the shortest between the dynamical time scale and the loss time scale. There are not many instances of non relativistic shocks that can lead to the generation of UHECRs. The most noteworthy exception is represented by shocks formed during the formation of clusters of galaxies \cite{jones}.

\begin{figure}
\begin{center}
\resizebox{.6\columnwidth}{!}{%
  \includegraphics{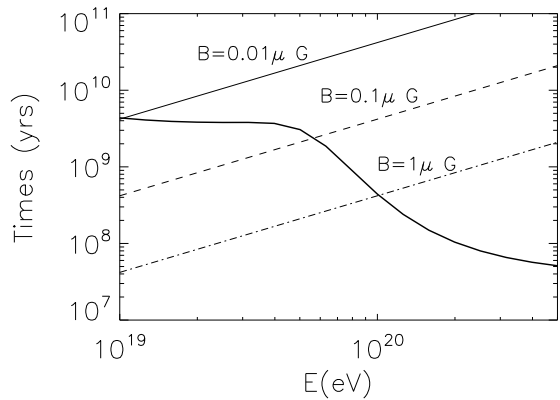} 
  }
\caption{Acceleration time for protons in accretion shocks moving at 5000 km/s with magnetic field $B_{0}=0.01$ (solid line), $0.1$ (dashed line) and $1 \mu G$ (dotted line), compared with the loss times due to interactions with the cosmic microwave background (thick line).  
}
\label{fig:clusters}       
\end{center}
\end{figure}

The formation of large scale structures in the universe has long been known to lead to the formation of supersonic plasma motions, and therefore to shock waves. Shocks can be formed during mergers of two clusters of galaxies or in the process of accretion of gas onto a gravitational potential well that already exists. This latter case applies to the filaments that connect clusters in the cosmic web. Typical velocities of several thousands of km/s are reached in these filaments, and the background temperature is expected to be relatively low ($\sim 10^{5}$ K), so that the shock waves that form are very strong with Mach numbers in the hundreds. Merger shocks, by comparison, are very weak since they develop in the intracluster medium which is already virialized ($T\sim 10^{8}$ K) \cite{gabici1}. The typical luminosity available at accretion shocks can be estimated as ${\cal L}\sim 10^{45}$ erg/s, in the right range to satisfy the constraints imposed in the section above. 

These accelerators operate for times which are comparable with the age of the universe, and in fact CRs accelerated at these shocks are confined in the intacluster volume for cosmological times \cite{storage}. The maximum energy is limited by energy losses. In Fig. \ref{fig:clusters} we show the loss time of protons (thick solid line) and the acceleration time at accretion shocks for different values of the magnetic field ($B=0.01$ (solid line), $0.1$ (dashed line) and $1 \mu G$ (dotted line)), assuming the optimistic case of Bohm diffusion. The most important feature that we notice is that for any reasonable choice of the magnetic field the acceleration time hits a wall of energy losses at $\sim 5\times 10^{19}$ eV. Cluster accretion shocks may represent an option as accelerators of UHECRs only if no substantial flux of UHECRs at $E>10^{20}$ eV is observed: the observed flux suppression would be an intrinsic cutoff in the accelerator but such a cutoff would be due to the same physical mechanisms responsible for the GZK feature. The spectrum accelerated at these shocks is very close to $E^{-2}$, thereby leaving little space to models of the transition such as the dip. 

\subsection{Relativistic shocks}\label{sec:relshock}

Our understanding of particle acceleration at relativistic shocks has still several unclear aspects. Qualitatively, the new ingredient is represented by the fact that both the particles being accelerated and the shock move relativistically. This reflects in the anisotropy of the distribution function of accelerated particles which has important implications on the return probability, energy gain and spectral shape. The first point to notice is that a relativistic shock is superluminal for all orientations of the background magnetic field that form an angle $>1/\Gamma$ with the shock normal. For large values of the shock Lorentz factor $\Gamma$, it becomes hard to avoid this condition. The consequences are quite important for shock acceleration: for $\Gamma\gg 1$, the shock velocity in the frame comoving with the downstream plasma is $\sim c/3$. On average a particle takes a time $\tau = 2\pi r_{L}/c$ to cover one Larmor rotation. In this time the shock moves by $\tau c/3=(2\pi/3)r_{L}>r_{L}$, namely the particle is trapped downstream and its probability of returning upstream is greatly reduced. This fact leads to expect steeper spectra for acceleration at relativistic shocks, as discussed in \cite{revenu}.

One may expect that large turbulence downstream may reduce this effect, leading to an increase in the return probability and to harder spectra. This assumption was implicitly made in all classical investigations on relativistic shock acceleration \cite{kirk,gallant,vietri1,vietri2} and in fact these calculations invariably led to expect a sort of universal spectrum $N(E)\sim E^{-2.3}$ when the additional assumption of small pitch angle scattering (SPAS) was made. In \cite{vietri2,ellison} the authors also explore the possibility to break the SPAS assumption, and find that the spectra can become appreciably harder than the universal spectrum quoted above. 

We will comment later on how realistic is to have large turbulence downstream. In the assumption that particles can make their way back upstream from the downstream region, it is worth asking what is the typical energy gain. The first time that a particle crosses the shock from upstream to downstream and back, its energy can increase by $\sim 4\Gamma^{2}$. For large values of $\Gamma$ this can be a sizeable energy gain: for instance for $\Gamma=300$ which can be achieved in GRBs, the first interaction between particles and the shock leads to particles with energy $\sim 3\times 10^{5}$ GeV, which becomes a low energy cutoff in the spectrum of accelerated particles. After the first shock crossing, particles are beamed within an angle $1/\Gamma$ around the shock normal. This strong anisotropy in the distribution function leads to an energy gain in any future shock crossing,  if they occur, of the order of $\Delta E/E\sim 2$. These simple arguments might become somewhat harder to make for the case of non-planar relativistic shocks, for instance for the case of relativistically moving plasmoids such as the ones that are observed sometimes in radio observations of jets arising from active galactic nuclei. 

Let us now move back to some considerations about the role of turbulent magnetic fields in relativistic shocks. Magnetic field can be amplified downstream of the shock and lead to enhanced scattering and to the return of particles from downstream. On the other hand, if turbulence exists or is excited upstream of the shock (for instance by CRs themselves) and eventually advected downstream, the situation is more complex. The perpendicular components of the magnetic field are compressed by a factor $\Gamma$ at the shock surface and the behaviour of particles at this point is determined by whether the scale of the turbulence is larger or smaller than the gyro-radius $r_{L}$ of particles in the field. If the magnetic field scale is smaller than $r_{L}$ one can expect the particles to scatter on a small scale turbulence. Return to upstream may be effective but the scattering is slow because of lack of resonance, therefore in this case the spectrum of accelerated particles may be the canonical one but the maximum energy is expected to be low. On the other hand, if the scale of the field is larger than $r_{L}$, then the shock behaves as perpendicular from the point of view of accelerating particles and the return probability is reduced, thereby leading to steep spectra. 

In conclusion, while relativistic shock acceleration has often been invoked as a mechanism for the acceleration of UHECRs, especially in the context of GBRs, there are many difficulties and poorly understood aspects of the acceleration process that do not allow us to reach a firm conclusion on whether this mechanism may be at work in potential UHECR sources such as AGN and GRBs. 

\section{Acceleration inspired by unipolar induction}
\label{sec:ns}

The rotation of a magnetized star leads to potentially large induced electric fields which may be responsible for particle acceleration \cite{bible}. This mechanism is often invoked in connection with black hole magnetospheres, and with fast spinning neutron stars. Here we consider the case of newly born neutron stars as possible sources of UHECRs. The original model was proposed in Refs. \cite{epstein} and \cite{arons}: the former version was concentrating on accelerated iron nuclei in Galactic neutron stars, while the latter paper focused on protons accelerated in extragalactic neutrons stars. 

The electric field induced by the rotating magnetic field dominates over the gravitational field and may extract electrons from the star surface. These electrons find themselves in the strong dipolar magnetic field of the star and suffer curvature radiation, which results in emission of photons, which in turn can pair produce by scattering with virtual photons associated with the magnetic field, and so on. This chain of events leads to a multiplication of the number of electrons and positrons that eventually fill the magnetosphere of the pulsar. The typical multiplicity (number of pairs generated by a single electron stripped off the surface) predicted by the models is $\sim 10-10^{4}$ depending on local conditions. The total potential drop that can be used for acceleration is:
\begin{equation}
V=\frac{\Omega R_{L}}{c} B_{s} \left( \frac{R_{s}}{R_{L}}\right)^{3} R_{L} = \frac{R_{s}^{3} B_{s} \Omega^{2}}{c^{3}} = 9\times 10^{20}  \left( \frac{\Omega}{3000 s^{-1}}\right)^{2} \left( \frac{B_{s}}{10^{13} G}\right) V
\end{equation}

The charge unbalance remains anchored to the so-called Goldreich-Julian (GJ) density \cite{goldreich}. The possibility exists that in addition to the extraction of electrons, some nuclei may be stripped off the surface of the star (on the opposite polar cap), although how realistic this possibility is depends on the very poorly understood structure of the crust of the star. If the crust has a lattice structure, electrons are easier to extract, because not tightly connected to the structure of the lattice, but in order to extract a nucleus the electric field must exceed the binding energy of the nucleus in the lattice. If nuclei are extracted, their density is expected to be of the order of the GJ density since they do not go through any cascading process typical of electrons. 

The fate of nuclei in the neutron star magnetosphere has been subject of much speculation: if they are energized by the induced electric field within the light cylinder, curvature energy losses are exceedingly large and the maximum Lorentz factor they can achieve is of no relevance for UHECRs \cite{bible}. One can invoke some different ideas so that nuclei acquire their Lorentz factor outside the light cylinder (see \cite{arons}) but this point remains rather fuzzy at the present time. As an order of magnitude the Lorentz factor of nuclei can be estimated as the typical magnetic energy at the light cylinder radius $R_{L}$, divided by the GJ density: $\Gamma=B^{2}(R_{L})/(8\pi n_{GJ} A m_{p} c^{2})$, corresponding to an energy: 
\begin{equation}
E_{max}(t) = 10^{21} \left( \frac{\Omega}{3000 s^{-1}}\right)^{2} \left( \frac{B_{s}}{10^{13} G}\right) eV
\end{equation}
for an Iron nucleus ($A=56$). Following \cite{epstein} we can estimate the spectrum that develops as a consequence of the spin down of the neutron star rotation, $N(E)\propto E^{\frac{1-n}{2}}$, where $n$ is the braking index ($n=3$ for a magnetic dipole). For a braking index $n\leq 3$ one can see that the spectrum is expected to be very hard (see also \cite{arons}). 

Many questions associated with this model remain open: 1) Can nuclei really be stripped off the surface of the star, namely is the electric field strong enough to win against the binding energy of nuclei in the lattice? 2) Once extracted, what fraction of the total potential drop is really accessible to nuclei, and where should the energization take place to avoid energy losses? 3) How do nuclei escape from the acceleration region? 

Very little can be done to address the first issue in any reliable way. It is interesting however that the presence of protons (or more in general nuclei) is required by a model of particle acceleration at the termination shock of pulsar wind nebulae (PWN) \cite{amatoarons}: although often overlooked, the acceleration mechanism that is responsible for the formation of non-thermal emission from PWN is all but understood. The standard Fermi acceleration at the relativistic termination shock does not work (see \S \ref{sec:relshock}). Even more so in the case of PWN, because the shock is dominated by electrons and positrons. In Ref. \cite{amatoarons} the authors proposed that acceleration of electrons and positrons may occur because of the resonant absorption of cyclotron waves produced by protons, which are allowed to exist in the relativistic wind of pulsars to the extent that their density does not exceed $m_{e}/m_{p}$ times the density of pairs (which is affected by the multiplicity of cascading in the magnetosphere, as discussed above). Should this model receive independent confirmation, it would represent a step forward in supporting the possibility that protons and nuclei may be accelerated in the pulsar environment. Some interesting new insights into the issue of escape of accelerated nuclei from the PWN, in the context of the model illustrated in this section have recently been discussed in Ref. \cite{fang}, who also summarize the theoretical ideas proposed so far to accelerate nuclei to very high energies avoiding energy losses. The authors point out two elements which may be very important for the model: 1) despite the very flat injection spectrum, the spectrum of particles that escape the host remnant is much softer as a result of spallation of iron nuclei on their way out of the remnant; 2) the same phenomenon also changes the composition of escaping particles with respect to the pure iron composition injected at the surface, making it mixed and potentially similar to the one observed by the Pierre Auger Observatory. 

\section{Gamma rays from UHECRs}
\label{sec:cascade}

Gamma ray observations can provide useful information on the origin of UHECRs. Gamma radiation is produced as a by-product of propagation of UHECRs on cosmological distances \cite{smirnov} and in fact the so-called cascade upper limit is often used to check whether a given model for the origin of UHECRs is viable. The limit is obtained by taking into account that Bethe-Heitler pair production and photopion production (with the respective kinetic thresholds) lead to energy conversion from cosmic rays to photons. Eventually the spectrum of photons in the cascade approaches a sort of universal spectrum \cite{smirnov}. Comparison of such spectrum with the observed diffuse isotropic gamma ray background allows one to impose limits on specific models of production of UHECRs \cite{berekach}. The cascade is mainly driven by two processes, pair production ($\gamma + \gamma_{EBL}\to e^{+}+e^{-}$) and inverse Compton scattering ($e^{\pm}+\gamma_{EBL}\to e^{\pm} + \gamma$), initiated by either an electron/positron produced in the Bethe-Heitler process or charged pion decay (only for positrons), or a gamma ray from the decay of neutral pions generated in $p\gamma_{EBL}$ reactions. The presence of magnetic fields may change the simple development of the cascade in that synchrotron losses result in the generation of very soft photons that do not take part in the development of the cascade. In all respects the synchrotron energy leaks out of the cascade. In this sense the presence of magnetic fields weakens the strength of the cascade upper limits. 

In addition to the diffuse gamma ray flux it is also worth considering the gamma ray flux due to cascading induced by UHECRs from the direction of individual sources of UHECRs. This idea was first investigated in \cite{ferrigno} for the case of nearby sources. At the time of that paper these local sources were expected to be responsible for the multiplets of events detected by the AGASA experiment. On such small distances ($\sim 100$ Mpc) the main channel responsible for the injection of gamma rays and $e^{\pm}$ is photopion production, while for distant UHECR sources Bethe-Heitler pair production becomes dominant despite its smaller inelasticity. Clearly the identification of the cascade flux as point-like requires that the magnetic field is very weak, in order to avoid spreading of the $e^{\pm}$ on too large angles, which would translate into diffuse radiation. 

This scenario was investigated in detail in \cite{kusenko}, with the purpose of explaining the gamma ray spectra of some distant AGN. The interesting finding is that the presence of weak magnetic fields, while not hindering the development of cascades, may cause the low energy gamma ray flux to spread in angle thereby losing the angular correlation with the point source. This results in a flux reduction in the GeV energy range. On the other hand, since CRs travel some distance from the source before suffering a $p\gamma_{EBL}$ interaction, photons may be injected closer to the Earth and there initiate a cascade, as described above. This effect may cause some very high energy gamma rays to come from the same direction as the source provided the magnetic field is not strong enough to deflect neither the parent proton nor the generations of $e^{\pm}$ produced in the cascade. This interesting effect was recently invoked as a possible explanation of the high energy gamma ray flux from a source with alleged redshift $z=1.2$ \cite{aharonian}. This redshift appears however to be very controversial and possibly incorrect. 

Although much caution need to be used for the interpretation of these data, the possibility to {\it see} the sources of UHECRs not directly but rather through the secondary photons in the electromagnetic cascade they induce by propagating in the EBL over cosmological distances remains a very promising way to search for the sources of UHECRs, especially in the perspective of upcoming Cherenkov telescopes such as CTA. 

\section{Summary}
\label{sec:summary}

At this moment in time, the problem of the origin of UHECRs is genuinely observational/experimental in nature. We have an unprecedented wealth of data from several experiments that provide information on the spectrum, chemical composition and anisotropy. Yet, these pieces of observations are not all in agreement with each other, suggesting that known and/or unknown systematics may affect the interpretation of the data. On one hand, HiRes and Telescope Array claim the detection of the GZK feature. This spectral feature, due to the reaction of photopion production of protons in their journey from the sources to Earth, suggests that UHECRs are protons. This is consistent with the value of $E_{1/2}$ measured by HiRes \cite{hires}. The elongation rate as measured by HiRes is also claimed to be compatible with a pure proton composition at $E>10^{18}$ eV. On the other hand, the Pierre Auger Observatory has detected a flux reduction at energies somewhat lower than the one claimed by HiRes. The chemical composition measured by Auger by using the elongation rate of its RMS fluctuations, appears to be mixed and dominated by heavy nuclei at the highest energies. In this case the detected flux reduction can hardly be the GZK feature and may either be modeled as the result of photodisintegration of nuclei or as an intrinsic cutoff in the source spectrum (see contributions by R. Aloisio and V. Berezinsky in this conference). Moreover, the Pierre Auger Collaboration has also measured a correlation of the arrival directions with the position in the sky of AGN in selected catalogs \cite{science}, although the statistical significance of such correlation weakened with time, to reach now a stable value of $\sim 2 \sigma$. The correlation is clearly explained more simply with protons as primaries, given the strong role of the Galactic magnetic field in deflecting particles. On the other hand, it is quite possible that what Auger is measuring is a proxy for a global anisotropy of the highest energy cosmic rays deriving from the fact that the sources (whatever they are) are concentrated where most local matter lies (the local supercluster). In this case, it is easier to understand that even heavier nuclei might retain some level of memory of the production region. In this sense, even from a semantic point of view, it might be more appropriate to talk about anisotropy of UHECRs rather than correlation with specific objects.

From the situation depicted above it is difficult to infer any solid conclusion about the nature of the flux reduction at $\sim 10^{20}$ eV, the chemical composition and therefore about the sources, not to mention the transition from Galactic to extragalactic cosmic rays. In this short review I have summarized some theoretical challenges that may be laid down even in the absence of clear information. 

Both models of extragalactic CRs that appear to be more appealing at the present time suggest that the transition from Galactic to extragalactic CRs occurs at energies below the ankle. The dip model explains the shape of the spectrum in terms of propagation of protons, and the ankle feature is simply explained as the result of a balance between the adiabatic losses due to the expansion of the universe and Bethe-Heitler pair production \cite{dip}. The model fits extremely well the modification factors as measured by all experiments with the possible exception of the one measured by the Auger Observatory. Moreover, ones the dip is interpreted as a result of a well known particle physics process (pair production), its position in energy is fixed. If this energy is considered as an absolute energy scale, and all energies in different experiments are normalized to the dip energy scale, the spectra also agree extremely well with each other, again with the exception of the Auger spectrum. The dip model works as long as not more than 15\% of the flux is contributed by He nuclei. 
  
Larger abundances of nuclei heavier than Hydrogen are described in the context of the mixed composition model. This model has more free parameters in the form of spectra and relative abundances of different nuclei. The overall spectrum and chemical composition as measured by Auger can be best fit in the {\it disappointing model} version, where the maximum energy of Iron is low and coincides with the flux suppression region as observed by Auger. 

The low energy of the transition from Galactic to extragalactic CRs, predicted in both the dip and mixed composition models, is appealing in that it appears to be in qualitative agreement with the expectation based on the supernova remnant paradigm for Galactic CRs.

As far as the sources are concerned, although gamma ray bursts and AGN appear as likely candidates, no specific reason exists to prefer one to the other and no reason exists to exclude other options such as rapidly rotating neutron stars. Again, the key to solve the problem of the origin of UHECRs is purely observational at this point: a careful and unambiguous measurement of the chemical composition and of the global anisotropy are crucial to move some steps forward in this field.

\end{document}